# Chiral, Topological, and Knotted Colloids in Liquid Crystals

Ye Yuan [1,*] and Ivan I. Smalyukh [1,2,3,*]

1. International Institute for Sustainability with Knotted Chiral Meta Matter (WPI-SKCM²), Hiroshima University, Higashi-Hiroshima, Hiroshima, 739-8526, Japan
2. Department of Physics, Department of Electrical, Computer and Energy Engineering, Materials Science and Engineering Program and Soft Materials Research Center, University of Colorado, Boulder, CO 80309, USA
3. Renewable and Sustainable Energy Institute, National Renewable Energy Laboratory and University of Colorado, Boulder, CO 80309, USA
* Correspondence: yeyuan@hiroshima-u.ac.jp (Y.Y.); iisinboulder@gmail.com (I.I.S.)

**Abstract:** The geometric shape, symmetry, and topology of colloidal particles often allow for controlling colloidal phase behavior and physical properties of these soft matter systems. In liquid crystalline dispersions, colloidal particles with low symmetry and nontrivial topology of surface confinement are of particular interest, including surfaces shaped as handlebodies, spirals, knots, multi-component links, and so on. These types of colloidal surfaces induce topologically nontrivial three-dimensional director field configurations and topological defects. Director switching by electric fields, laser tweezing of defects, and local photo-thermal melting of the liquid crystal host medium promote transformations among many stable and metastable particle-induced director configurations that can be revealed by means of direct label-free three-dimensional nonlinear optical imaging. The interplay between topologies of colloidal surfaces, director fields, and defects is found to show a number of unexpected features, such as knotting and linking of line defects, often uniquely arising from the nonpolar nature of the nematic director field. This review article highlights fascinating examples of new physical behavior arising from the interplay of nematic molecular order and both chiral symmetry and topology of colloidal inclusions within the nematic host. Furthermore, the article concludes with a brief discussion of how these findings may lay the groundwork for new types of topology-dictated self-assembly in soft condensed matter leading to novel mesostructured composite materials, as well as for experimental insights into the pure-math aspects of low-dimensional topology.

## 1. Introduction

Liquid crystals (LCs) are characterized by the long-range orientational order of constituting mesogens with anisotropic shapes, with the simplest form being a nematic LC consisting of rodlike molecules [1–3] (Figure 1a). Despite the crystal-like anisotropic properties stemming from the orientational ordering, the weak intermolecular interactions between LC mesogens still allow for uninhibited translational motions of the building blocks and thus liquid-like flow, earning the name "liquid crystal" of the medium that combines properties of crystals and liquids. Different from isotropic fluids such as water, LCs feature large anisotropic dielectric and optical properties as a result of the molecular orientation field. The combination of such anisotropy and the facile responsiveness to external stimuli enables far-reaching technological applications, notably the flat-panel liquid crystal displays. Such properties of LCs also provide a unique anisotropic environment for colloidal particles, giving rise to rich phenomena not observed in conventional isotropic solvents, opening avenues for addressing fundamental physics questions as well as exploring pre-designed development of novel functional materials [4–8]. Of particular interest is how the geometric shape, symmetry, and topology of colloidal particles interact with the LC molecular orientation field to enrich this behavior, leading to unusual topological field distributions, colloidal interactions, and self-assembly processes.

Colloids are mixtures where particles of 1 nm to 10 μm are stably suspended in a solvent [9–11]. Particles of this size range are susceptible to Brownian motions which ensure the equilibrium suspension of colloidal systems, and are also responsible to other unique colloidal behaviors including light scattering, depletion interaction, capillary assembly, etc. Common everyday objects including milk, fog, and clay are all colloids in nature, while artificial colloidal systems have found applications in cosmetics, energy harvesting, smart materials, etc., as well as in understanding fundamental science questions ranging from mechanisms of phase transitions to emergent properties of self-assembled materials [9–14]. When introduced to a LC medium, colloidal particles force the surrounding LC molecules to reorient in order to accommodate the volume occupied and interfacial boundary conditions imposed by the foreign objects. Such accommodation manifests as the redistribution of the LC orientation field including both continuous deformations and singular defects, which can spatially propagate far beyond the physical extent of the colloidal particle. The equilibrium state of the LC colloidal system is determined by the configuration that minimizes the total energy cost associated with the field distortions, which also leads to long-range colloidal interactions mediated by the LC anisotropic elasticity. With the ability to define how the LC orientation field is disturbed, the size, shape, and surface properties of colloidal inclusions become important contributing factors determining the equilibrium states and dynamic processes of the composite LC colloidal systems.

Early studies of LC colloids were usually carried out on highly symmetric particles of trivial topological characteristics, such as spheres, rods, or discs [15,16]. The technological advancement in colloidal fabrication [4,5,17,18] and microscopic imaging [19] in recent years has allowed researchers to obtain colloidal particles of much more complex morphological characteristics, as well as resolve fine details on LC field distributions induced by these particles. Examples of fabrication techniques include two-photon photopolymerization (2PP) and new chemical synthesis protocols to obtain chiral or knotted colloidal particles in nano and micrometer size ranges and made of a large range of constituent materials ranging from polymerizable resin to noble metals, which are reviewed elsewhere [18]. Imaging and optical manipulation techniques include three-photon excitation fluorescence polarizing microscopy (3PEF-PM) and holographic laser tweezer. Concurrently, developments in computational modeling have enabled accurate simulations of complex LC colloidal systems, providing insights into the underlying physics and guiding experimental design [4,5,20].

In this review, the authors intend to provide a concise account of nematic LC colloids containing particles of complex geometries such as chiral, topological, knotted, and linked features. These types of colloidal particles impart their symmetry-breaking and topologically nontrivial geometries into the LC orientation field in the forms of chirality-dictated interactions, topological constraints on singular defects, as well as knotted field configurations, which is of both fundamental science and application significance. The review is structured as follows: after these introductory statements and remarks, the following section will lay out the physical underpinnings of colloidal particles in LCs, overviewing the relevant theories of LC elastic energy and topological defects, which will be followed by three main sections discussing colloidal particles of various symmetry and topology characteristics; in the last section, we discuss further the context and implication of LC colloids and potential future directions of development for this field.

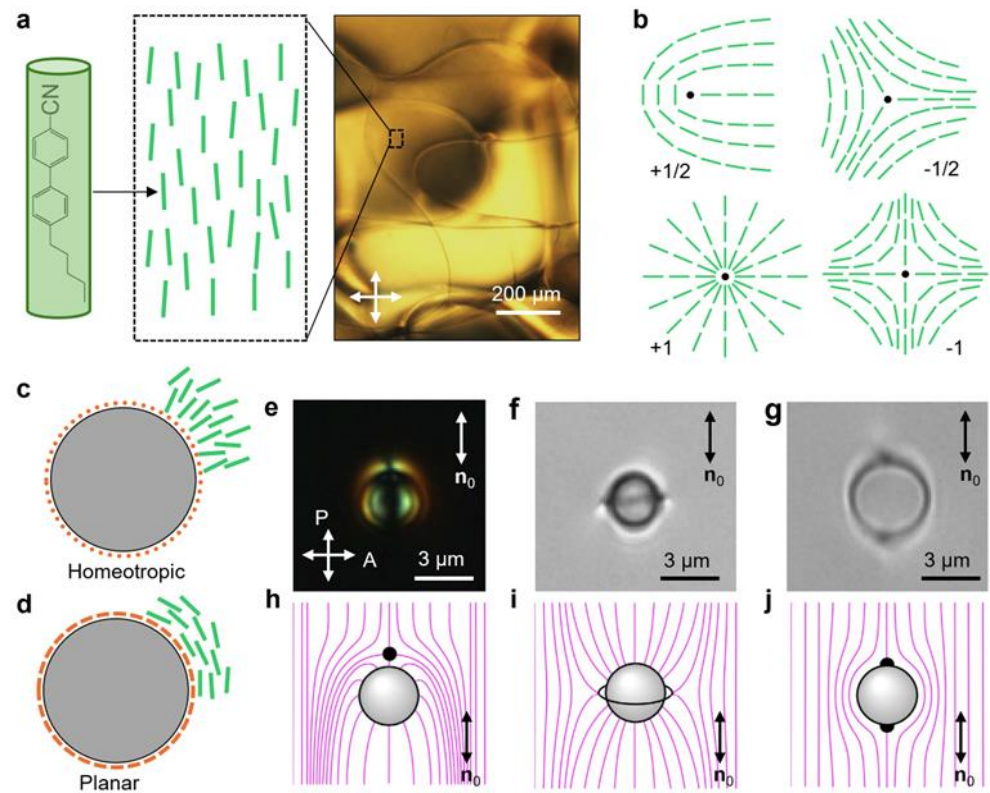

**Figure 1.** Colloids in liquid crystals (LCs). (**a**) Microscopic structure of a nematic LC with rod-like mesogens, i.e., pentylcyanobiphenyl (5CB). The micrograph (right image) shows the texture of a 5CB droplet observed under a microscope with crossed polarizers, polarization direction marked with white double arrows. Inset shows the chemical structure of 5CB molecules and their collective alignment within a small volume. (**b**) Topological defects in LCs of different winding numbers; green rods represent LC molecules. (**c**,**d**) Homeotropic and planar surface anchoring where LC molecules align perpendicular and parallel to the surface of colloidal inclusions. Orange dots and dashes represent surface functioning agents such as polymer grafting that impose the anchoring direction. (**e**,**f**) Micrograhs showing microspheres with homeotropic surface anchoring inducing "hedgehog" point defect and "Saturn ring" line defect; white double arrows indicate the crossed polarizers. (**g**) Micrographs showing microsphere with planar surface anchoring inducing "boojum" surface defects at the polar points of the sphere. (**h–j**) Corresponding schematics illustrating LC director field configurations around the colloidal spheres. The black dots and line represent the hedgehog defect, the Saturn ring loop, and the surface boojums, respectively. Schematics are not drawn to scale. The far-field director is shown by the double arrow marked with **n**$_0$. Adapted from Ref. [21].

## 2. Basics of Physical Behavior of Colloidal Particles in LCs

The LC molecular ordering can be described using the so-called director field **n**(**r**) representing the local average of the molecular alignment, which can vary as a function of coordinates **r** on scales much larger than molecular dimensions. In contrast to a vector field, the constraint **n**(**r**) = −**n**(**r**) is imposed to comply with the non-polar nature of the LC molecular orientations [1]. In the undisturbed state, molecules of a nematic LC can orient along the same direction with the help of proper surface confinement, typically setting surface boundary conditions corresponding to a uniform distribution of the director **n**(**r**) = **n**$_0$, which also corresponds to the energy-minimizing state of a uniaxial nematic LC. Deviating from such a "ground state" uniform alignment via producing spatial gradients of the director is energetically costly, which can be quantitatively described using the Frank-Oseen free energy density functional $f_e$ [1]:

$$f_e = \tfrac{1}{2}K_{11}(\nabla \cdot \boldsymbol{n})^2 + \tfrac{1}{2}K_{22}(\boldsymbol{n} \cdot \nabla \times \boldsymbol{n})^2 + \tfrac{1}{2}K_{33}(\boldsymbol{n} \times \nabla \times \boldsymbol{n})^2 - K_{24}\nabla \cdot [\boldsymbol{n}(\nabla \cdot \boldsymbol{n}) + \boldsymbol{n} \times (\nabla \times \boldsymbol{n})], \qquad (1)$$

where $K_{11}$, $K_{22}$, $K_{33}$, and $K_{24}$ are the so-called elastic constants characterizing the energetic costs of splay, twist, bend and saddle-splay director deformations. The typical value of elastic constants is a few pN, for example for cyanopentylbiphenyl (5CB), a very common LC that is in the nematic phase at room temperature (Figure 1a). In colloidal systems, where the characteristic length scale of colloidal inclusions can be $L \sim 1$ μm, the ratio between the elastic and the thermal energy is $KL/k_BT \sim 10^3$ ($K \sim 1$–10 pN is a representative elastic constant of LCs, $k_B$ is the Boltzmann constant and $T = 300$ K is the room temperature). Therefore, the elastic energy is sufficient to overcome thermal fluctuations and lead to elasticity-mediated colloidal interactions and self-assembled structures in LCs.

Another form of energetic cost is related to the surface of the colloidal particles. Boundary conditions imposed by grafted polymer or other functional groups can also define a preferred direction (often called "easy axis") for the LC molecules to align to, which can be normal, tangential, or sometimes conical to the surface [15,16,21]. The two most common cases are homeotropic (normal) and planar (tangential) surface anchoring (Figure 1c,d). The free energy density functional characterizing the energy cost of the director deviation from the easy axes takes the Rapini-Papoular form [1]:

$$f_s = -\frac{1}{2}W(\boldsymbol{n} \cdot \boldsymbol{e})^2 \qquad (2)$$

where $\boldsymbol{e}$ is a unit vector along the easy axis and $W$ is the surface anchoring strength with a typical value in the range $10^{-6}$–$10^{-4}$ J/m²; integrating this free energy density over the area of LC interfaces with particles or confining substrates gives the overall energy cost of LC-surface interaction. A parameter called "extrapolation length" can be defined as $\xi_e = K/W$, yielding typical values in the range $\sim 10^{-2}$–1 μm. When particle's size is much smaller than $\xi_e$, it cannot disturb the director field significantly enough and thus the bulk elastic energy contribution can be ignored; conversely, in the opposite regime where the boundary conditions defined by the easy axis orientation at the interface can be assumed infinitely strong, the energy cost of inserting the particle into the LC consists of only that from elastic deformations and singular defects around the particle. Overall, in the most general case, the equilibrium distribution of the director field around the colloidal inclusions are determined by minimizing the total free energy of the system, which includes integrating over the bulk volume of the system $V$ and the enclosing surface $S$ for the bulk and surface energies, respectively:

$$F_{\text{total}} = \int_V f_e \, dV + \int_S f_s \, dS. \qquad (3)$$

When multiple particles are present, however, they can move relative to each other to minimize the total energy, which effectively determine the elasticity-mediated colloidal interaction patterns in LCs. Under the one-constant approximation where all elastic constants are equal in value, the director field with small perturbations far away from the colloidal particles $\mathbf{n} = (n_x, n_y, 1)$ follows the Laplace-like equation [22]:

$$\Delta n_\mu = 0 \; (\mu = x, y) \qquad (4)$$

obtained by minimizing the Frank-Oseen free energy. Like that in electromagnetic theories, this equation can be solved using multipole expansion expressed in terms of spherical harmonics, and the colloidal interactions mediated by the LC elasticity can thus be interpreted in the multipole paradigm [22–25]. However, it is worth mentioning that in the regime when colloidal particles are too close to each other, such that the small-perturbation approximation may no longer hold true, contributions from higher-order multipoles and nonlinear field effects may appear, which goes beyond the multipole expansion analysis. While the first experimental demonstrations of elastic interactions resembling that between electrostatic multipoles were made in 1997 when Poulin et al.

discovered dipole–dipole-like interactions between water droplets with homeotropic anchoring in a nematic LC [15], Brochard and de Gennes in 1970 had considered such electrostatic analogy theoretically [22]. Thus far, elastic multipoles of zeroth till fourth order (monopole to hexadecapole) have been experimentally demonstrated using various type of LC colloidal systems [15,21,26,27] (Figure 1d), paving the way for establishing LC colloidal systems with structures and emergent behaviors mimicking that of atomic systems [28].

In addition to continuous deformations, singularities where the local director cannot be well-defined may also be induced by colloidal inclusions [3]. These singularities can exist in the bulk as defect points (also called hedgehogs) and defect lines (disclinations), as well as on the surface of the colloidal particles, where the surface point singularities are called boojums. In fact, the name "nematic" originated from the Greek word meaning thread, refers to the thread-like disclination lines observed in a nematic LC (Figure 1a). The defects can be classified by computing the winding number in 2D for surface boojums and cross section of disclinations, or topological charge in 3D for hedgehogs and compact closed loops of disclinations [29,30]. The winding number $s$ counts the number of times as the director rotates by $2\pi$ following a loop circumnavigating the defect core once (Figure 1b); the sign indicates whether the rotation direction of the director is the same or opposite to that of the circumnavigation. Similarly, the topological hedgehog charge $m$ counts the number of times the unit sphere is wrapped by the director on a surface enclosing the defect core [30]:

$$m = \frac{1}{4\pi}\int_S \ (\boldsymbol{n} \cdot \partial_1 \boldsymbol{n} \times \partial_2 \boldsymbol{n}) \, dx_1 dx_2. \tag{5}$$

Several examples of singular defects with winding number of ±1/2 and ±1 are illustrated in Figure 1b. In addition to representing the director distribution of surface boojums, they also serve as the cross-sectional schematic of disclination lines, with the defect core extending out of the viewing plane.

The topology of colloidal particles imposes constraints on the generation of singular defects. Following the Gauss-Bonnet theorem and assuming uniform far-field director $\mathbf{n}(|\mathbf{r}| \gg L) = \mathbf{n}_0$, one finds that the total topological charge induced by a colloidal particle compensates that due to its boundary conditions and is equal to half of particle surface's Euler characteristic [31]

$$\sum_i m_i = \pm \chi/2. \tag{6}$$

The Euler characteristic $\chi$ is a topological invariant that can be computed, for example, based on the number of vertices, edges, and faces of the surface, denoted by $V$, $E$, and $F$, respectively

$$\chi = V - E + F. \tag{7}$$

For surface boojums in cases of tangential degenerate surface boundary conditions, the mathematical theorems impose the requirement on the total winding number of surface defects [32]:

$$\sum_i s_i = \chi. \tag{8}$$

However, it is important to note that the topological constraints alone cannot determine the exact distribution of topological defects. The constraint limits the net topological charge or winding number, while it is the minimization of the total free energy that determines the director field deformation and defect distribution that satisfy this topological constraint. For example, the elastic energy per unit length of a disclination line scales with the winding number as $\propto s^2$ [3], indicating that disclinations of higher strength involve significant field distortions and, thus, are energetically unfavorable, tending to split into lower-winding-number ones. For colloidal spheres with a homeotropic surface anchoring, either a hedgehog defect or a disclination loop named "Saturn ring" can be found accompanying the particle (Figure 1e,f,h,i). While both configurations satisfy the constraint in Equation (6) since spherical surfaces have an Euler characteristic of 2, the Saturn ring is the energetically preferred one when the sphere is small, i.e., <1 μm [29]. In contrast, the constraint in Equation (8) is fulfilled by two +1 surface boojums at both poles for a colloidal sphere with planar surface anchoring (Figure 1g,j). More discussions on the interplay between topological defects and colloidal particles of non-trivial topology are in Section 4.

In addition to the literature referenced above, interested readers can refer to other in-depth reviews and textbooks [3–5].

### 3. Chiral Colloids

Chirality manifests itself in many different scales, ranging from subatomic elementary particles to molecular enantiomers with opposite optical activity and to biological and even cosmological objects [33]. Chiral colloidal particles suspended in a LC may transfer the broken symmetry manifestations into the surrounding medium, for example, in the form of low-symmetry director field distributions both in the close vicinity of the particle and in the far field, leading to topological defect distributions and novel colloidal interactions mediated by the LC elasticity. This is in direct contrast to colloidal spheres in nematic LCs inducing director distortions resembling symmetric distributions corresponding to dipoles, quadruples, or higher-order multipoles [5,16]. Martinez et al. used 2PP-based 3D microprinting to fabricate custom-designed chiral microparticles bound to substrates in LC cells (Figure 2a,b) [34]. Nonlinear optical imaging reveals that these surface-attached particles impose a twist on the LC director field that propagates across the cell and generates low-symmetry director distortions. A colloidal sphere inducing elastic dipole moments is found to be attracted to the surface-bound particles following monopole–dipole interaction scaling (Figure 2c). By engaging surface structures with chiral features, this finding provides a new way to control LC molecular alignment and the ensuing elastic interactions, demonstrating possibilities of directing the self-assembly of colloidal particles through engineered surfaces.

The fundamental role of particle chirality on colloidal behaviors is further revealed with free-standing colloidal structures with chiral symmetry. Yuan et al. fabricated colloidal springs and helices of both handedness and re-suspended them in a nematic LC (Figure 2d–i) [35]. Despite their complex shape, these chiral structures are topologically equivalent to spheres, with an Euler characteristic $\chi$ = 2. Given their planar surface anchoring, the winding numbers of the surface boojums should sum up to two following Equation (8). For the colloidal springs, two stable (corresponding to the lowest energy states) and metastable (local energy minima) orientations are observed experimentally, with their helical axes tilting slightly away from being perfectly orthogonal and parallel to $\mathbf{n}_0$, respectively (Figure 2d,e). In the stable state, the topological constraint is satisfied by two $s$ = 1 boojums located at the edges of the spring's two end faces, while the metastable state has additional pairs of self-compensating boojums at positions where the surface normals are parallel to $\mathbf{n}_0$. Additionally, dipole-like pair interactions arise due to the director field deformations with chiral symmetry induced by the microsprings: the interaction is dependent on the angle that the center-to-center separation vector makes with $\mathbf{n}_0$ and the interaction potential scales as $d^{-3}$, as expected for dipole–dipole interactions (Figure 2h,i). The interesting observation is that the equivalent dipole direction depends on the relative handedness: same-handed microsprings interact as if they possess the same direction of dipolar moments, while the interaction direction flips when one of the springs changes its handedness (Figure 2f,g). These experimental observations are further supported by the numerical modeling based on the minimization of the Landau-de Gennes free energy. The shape anisotropy of such colloidal particles also brings in anisotropic diffusive behavior coupled with the LC director field [36]. Overall, both reports demonstrate how chirality can be used as an important parameter to control and engineer the colloidal interactions in LCs. The findings bridge the microscopic chirality of individual particles with mesoscale self-assemblies of colloidal structures, which may enable novel optical, photonic, and other functional materials. While chirality transfer from guest molecules to the host LC medium is widely known and technologically utilized, e.g., in cholesteric LC displays and electro–optic devices, the fact that colloidal inclusions with dimensions three orders of magnitude larger than the host LC's molecules can also transfer chiral symmetry breaking into the otherwise achiral nematic host is a fundamentally important finding with a significant technological potential from the standpoint of view of the self-assembly of mesostructured functional materials.

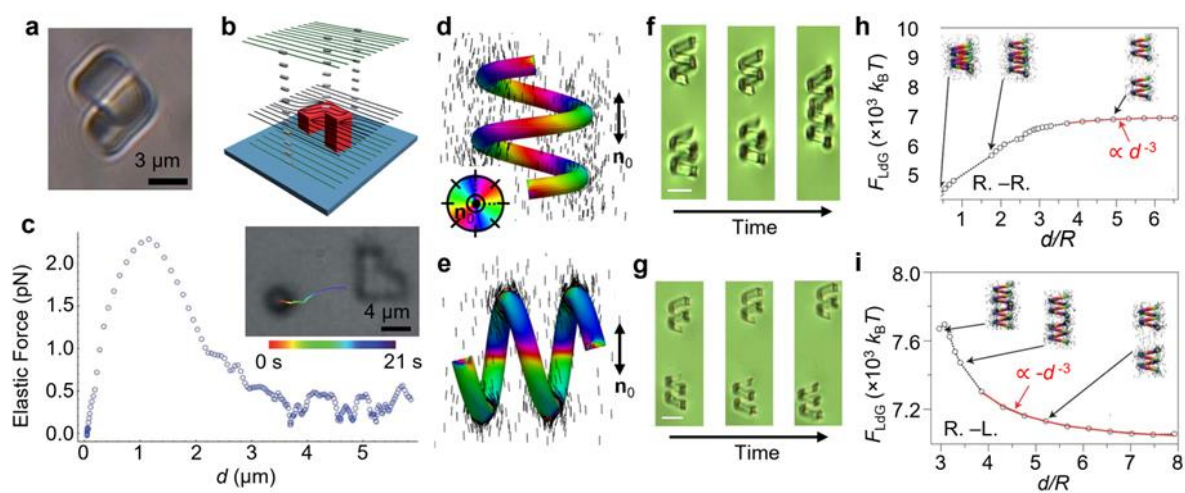

**Figure 2.** Chiral colloids in LCs. (**a**) Micrograph of a chiral microstructure obtained by 3D microprinting. (**b**) Director field distortions around such a particle with planar surface anchoring bound to the confining substrate. The lines in the middle layer and the cylinders show that twist deformation is induced in the director field over the particle despite uniform far-field alignment. (**c**) Interaction forces vs. distance between the surface-bound chiral structure and a free-floating colloidal sphere. The inset is a micrograph of the interacting objects and the interaction trajectory is color-coded with time. (**d**,**e**) Director distributions around right-handed microsprings with a planar surface anchoring at energy-minimized positions. The double arrows indicate the far-field director $\mathbf{n}_0$; the color on the particles represents the orientation of the surface director projected to the plane orthogonal to $\mathbf{n}_0$; color scheme is shown as the inset of (**d**). (**f**,**g**) Snapshots of elasticity-mediated interactions between like- (**f**) and opposite- (**g**) handed microsprings, exhibiting attraction and repulsion, respectively, over the time of 10–100 s. Scale bars are 5 μm. (**h**,**i**) Numerically calculated Landau-de Gennes free energy vs. particle distances between like- (**h**) and opposite- (**i**) handed microsprings. When particles are far away from each other, the free energy scales as $d^{-3}$ like that of dipole–dipole interactions. The distance between particles $d$ is normalized by the particle radius $R$; the free energy $F_{LdG}$ is normalized by the thermal energy $k_BT$, where $k_B$ is the Boltzmann constant and $T$ is the room temperature. Adapted from Refs. [34,35].

## 4. Topological Colloids

Although not explicitly emphasized, colloidal particles of non-trivial topology, i.e., non-homeomorphic to spheres (square particles with a hole at the center) were introduced into a LC as early as 2009 (Figure 3a,b) [37]. These square particles induced quadrupolar deformations in the director field, though the central hole did not play an important role in determining the colloidal interactions in this particular system, which was the focus of that study [37]. The importance of particle topology, however, lies in its ability to control the distribution of induced topological defects by defining the net topological invariant of the defects as described in Section 2. While the cases are simple for topologically trivial spherical particles, the most common ones in colloidal sciences, Senyuk et al. and Liu et al. fabricated colloidal particles of nontrivial topology, handlebodies of various genus (i.e., number of holes) (Figure 3c,d), and demonstrated that the total charge or winding number of topological defects induced by particles is consistent with topological surface characteristics of the colloidal inclusions, following the relations given by the Gausss–Bonnet and Poincare–Hopf index theorems [31,32]. For the surface of a particle, its Euler characteristic is related to the genus $g$ by $\chi = 2-2g$, where genus $g$ can be intuitively understood as the number of holes of the surface. Thus, the topological constraint in Equations (6) and (8) becomes [31]

$$\sum_i m_i = \pm(1-g), \tag{9}$$

and [32]

$$\sum_i s_i = 2-2g. \tag{10}$$

The relations were experimentally confirmed while $g$ was varied from 1 to 5, corresponding to handlebodies with 1 to 5 holes. For a $g$ = 1 handlebody particle with homeotropic surface anchoring, the most commonly observed orientation is the particle plane being perpendicular to $n_0$ to minimize the free energy cost. Curiously, although no defects are topologically required to form as Equation (9) is 0 when $g$ = 1, the constraint is satisfied by a pair of self-compensating disclination loop or an exterior disclination loop and a hedgehog defect at the particle center, totaling net zero on the sum of the topological charge (Figure 3c,d) [31]. The same type of handlebody particle with $g$ = 1 and planar surface anchoring induces two pairs of opposite-sign boojums so that the net winding number is 0 [32]. More interestingly, the topological constraints are still obeyed even when external stimuli cause significant redistribution of the director field. This emergent behavior shows how topological defects can nontrivially emerge while both satisfying pure math theorems and yielding energetic minima of the LC–colloidal soft matter system. Scenarios demonstrated in the above reports [31,32] include explorations of inter-transformations between point defects and disclination loops at the central regions of the handlebodies' holes when locally melting the LC, or when particles are reoriented following the application of external field. In other words, the total charge is a conserved quantity that is dictated only by the topological characteristic of the defect-inducing particles, i.e., the Euler characteristic of the 2D LC–colloidal interface.

It is an important result vividly demonstrating how topological theorems result in constraints on induced defects that are still "soft" in the sense of allowing the system to choose defect configurations of a certain net topological charge that are also energetically favorable. This energy-dependent selection of topology-satisfying configurations distantly resembles other types of director field transformations driven by energy minimization, such as the radial director configuration in a cylindrical homoetropic confinement escaping into the third dimension [38]. Moreover, variations in morphological features of colloidal particles can significantly diversify the type of defects and director deformations induced and modulated by the surface topology even within the same constraint. Below, we discuss two examples both involving $g$ = 1 particles, where we find that the ensuing defects and director distributions are drastically different because of differences in the details of the particle geometry. Senyuk et al. discovered quarter-strength defect lines pinned to the sharp edges of torus-like particles with a large cross section of the tube measuring 5 × 5 μm² (compared to 1 × 1 μm² of the handlebodies in Refs. [31,32]) (Figure 3e) [39]. While the total topological charge is still consistent with Equation (9), the quarter-strength defect lines are observed to migrate between edges of the particles by transforming into half-integer disclinations across the surface of the particle, causing the particle to tilt in its energy-minimizing orientation (Figure 3e–g). The disclinations can also be stretched and repositioned using optical tweezers while conserving the total topological charge. A more extreme case of changing geometric features while preserving topology involves fractal particles with many defect-inducing corners and edges once introduced to a LC. Hashemi et al. fabricated Koch star-based fractal particles from the 0th to the 3rd fractal iteration order [40]. As particles of genus $g = 1$ with homeotropic surface anchoring, the total topological charge induced should be 0 even though fractal shapes at high iteration orders exhibit complex geometric details (Figure 3h). It is observed that this constraint is fulfilled by pairs of oppositely charged defect points and disclinations found at the corners of the fractal particles, with the number of defects growing exponentially following the increased fractal order. However, when the fractal feature size is comparable to the nematic correlation length characterizing the mesoscopic size of the nematic ordering, director regions of reduced order parameters start to overlap, effectively causing local nematic-to-isotropic phase transitions in the vicinity of the fractal corners. Self-similarities of the nematic response such as defect distributions following the fractal stimulus are observed across different fractal iteration orders, despite being limited by the nematic correlation length. This work demonstrates that colloidal particles with fractal morphological complexity, while following the constraints imposed by the topological theorems, exhibit emergent behavior stemming from their fractal nature and may be instrumental in probing and understanding nematic order interactions with confining surfaces at limiting length scales.

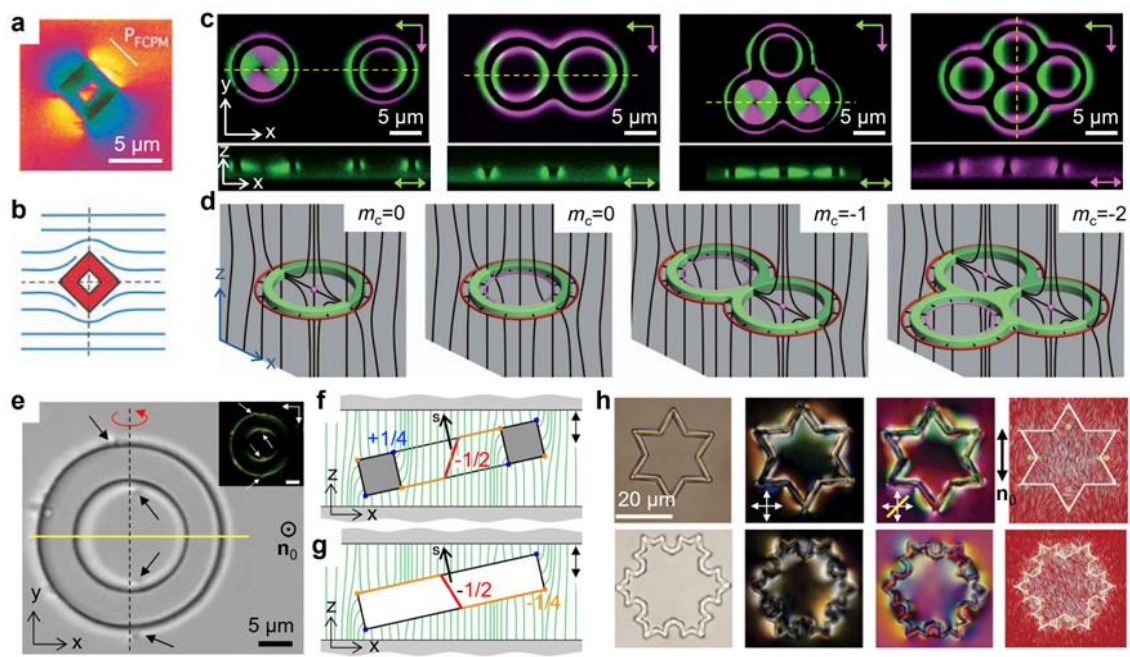

**Figure 3.** Topological colloids in LCs. (**a**,**b**) A square platelet with a central opening suspended in a LC. The image in (**a**) is taken with fluorescence confocal polarizing microscopy (FCPM); P$_{FCPM}$ indicates the polarization direction of the excitation light. The schematic in (**b**) shows the director distribution around the platelet. (**c**,**d**) Colloidal handlebodies of various genera in LCs. Panels in (**c**) are micrographs obtained by overlapping fluorescence images with orthogonal excitation polarizations as indicated by the green and magenta arrows; insects below are cross-sectional images in the *xz* plane taken along the yellow dashed lines. The schematics in (**d**) represent the director (black lines) distortions and topological defects (red and purple lines; purple dots) induced by the handlebodies; the total topological charge is determined by the particle genus $m_c = 1 - g$. (**e**–**g**) A large torus-shaped colloidal particle with homeotropic surface anchoring suspended in a nematic LC, inducing ½ and edge-pinned ¼ defect lines. The ¼ defect lines (blue lines in the schematic (**f**,**g**)) traverses along the edge of the particle and may jump between edges connected by ½ defect lines (red lines in (**f**,**g**)). The black arrows in the micrograph in (**e**) indicate the location of bulk ½ defect lines; tilting of the particle is indicated by the dashed line (rotation axis) and red curved arrow (tilt direction). Insets in (**e**) are obtained under crossed polarizers; polarization marked

by white arrows. (**h**) Fractal colloidal particles with homeotropic surface anchoring in LCs. The first column is taken at elevated temperatures when the surrounding LC is in isotropic phase; the middle two columns are taken under crossed polarizers (white double arrows) and crossed polarizers with a retardation waveplate (yellow line). The last column is a computer-simulated director field distribution around such particles. Adapted from Refs. [31,37,39,40].

In addition to the static distributions of topological defects and director deformations, dynamic processes involving topological colloids in LCs provide the possibility of exploring the out-of-equilibrium interplay between particle topology and the topological feature of the director field. One cannot change the topology of a particle's surface without breaking it. However, the effective topology can be changed, for example, by smoothly opening and closing holes in a 2D surface. Exploiting the anisotropic shape deformation of LC elastomers upon phase transitions, Yuan et al. fabricated ring-shaped colloidal particles and discovered that when heated, the aspect ratio of the rings changes and eventually, the closed holes at the center of the particles open up after the temperature is above the nematic–isotropic phase transition point (Figure 4a) [41]. Moreover, 2D crystal-like lattices of such LC elastomeric rings suspended in the unpolymerized nematic host show anisotropic collective actuations upon phase transitions. The experiments conducted so far, with the director fields of elastomeric particles and unpolymerized LC matching at the interfaces, do not explore the induction and transformation of defects as the net topological charge of the director inside and outside the rings stays neutral. However, how such a change in particle topology may lead to the evolution of topological features of the director field for cases of particles made of non-LC material and with well-defined boundary conditions at interfaces has been considered computationally. Numerical simulations of such systems by Ravnik et al. have provided insights: for a toroidal particle with two disclination loops, shrinking of the central hole results in the inner disclination looping closing into a point defect and eventually disappears as the hole is fully closed (Figure 4c) [36]. We may expect that future developments in the area of shape-morphing colloids may allow for the direct observation of the interaction and dynamic interplay between the topological characteristics of all components.

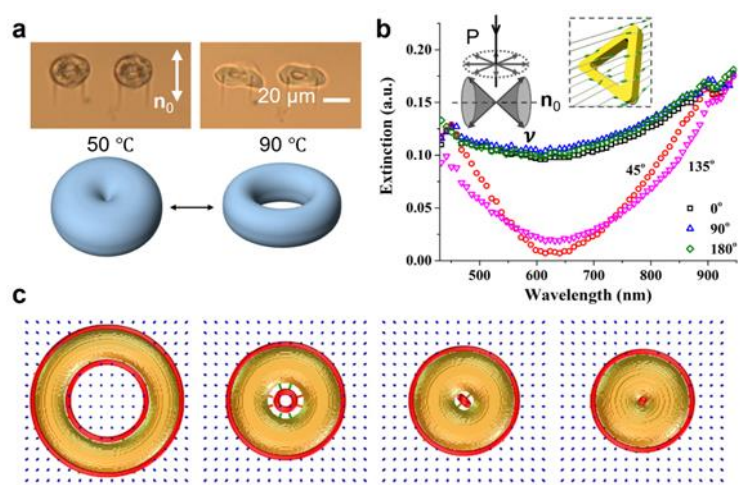

**Figure 4.** Stimuli-responsive topological colloids. (**a**) Ring-shaped microparticles made with liquid crystal elastomers change shape upon temperature elevation over the nematic–isotropic phase transition point. The schematic in (**a**) shows the opening and closing of the central hole, effectively changing the topology of the particle. (**b**) Alignment and polarization-dependent extinction of plasmonic triangular nanoframes dispersed in a nematic LC. The schematic in (**a**) shows the director distortions caused by the triangular frame. The normal $\nu$ of the plane containing the nanoframe has the freedom to rotate in a cone shape with the far-field director $n_0$ as the symmetry axis; **P** represents the polarization of the incident light. (**c**) Contraction of a ring-shaped particle with homeotropic anchoring and the induced disclination loops (indicated by red lines) in a nematic LC host. The far-field director is perpendicular to the viewing plane. Adapted from Refs. [41–43].

Another interesting possibility recently considered experimentally is the mesoscale material behavior under conditions when the plasmonic nanoparticles dispersed in a nematic host have nontrivial topology. The complex geometry of triangular nanoframes made of gold and silver enables the generation of plasmonic effects as well as the cone-like orientation distribution of the particles' normal relative to the director when suspended in a nematic LC [43]. Differently from micron-scale dielectric particles with similar topology [44], polarization direction-dependent extinction spectra are observed upon incident light in the visible and infrared range (Figure 4b), consistent with the symmetry prescribed by the nanoframes' orientation distribution in the LC. Furthermore, this guest-host system enables the electric switching of the plasmonic extinction of the hybrid LC-nanocolloid material reminiscent of that of pure LCs in terms of timescales and threshold voltages but now allowing for tuning the surface plasmon resonance effects. This work complements existing works on plasmonic LC colloids with trivial topology at the time of publishing and may now be extended to other plasmonic particles of broken symmetry or nontrivial topology [45].

## 5. Knotted and Linked Colloids

Different from our everyday experience, a knot commonly studied in topology cannot be undone because the ends are connected to form a closed loop. The simplest form of a mathematical knot is a ring, a.k.a., an unknot, the behavior of which as a physical object with a tubing-like surrounding immersed in LCs is described in the previous section as genus $g$ = 1 particles. More complex structures, i.e., non-trivial knots, in a LC director field induce knotted, linked, and other topologically non-trivial field configurations, allowing for experimental insights into predictions from knot theory and into the interplay of topologies of knotted surfaces, field deformation, and defects. Martinez et al. fabricated colloidal trefoil and pentafoil knots of tangential and homeotropic surface anchoring and demonstrated that a rich variety of director field and defect configurations may arise from the minimization of the total elastic energy of the system while still following the same topological constraints imposed by theorems from pure math (Figure 5a–e) [46]. For all cases studied of torus knots in the paper, the Euler characteristic is $\chi$ = 0, the same as for rings or $g$ = 1 handlebodies, which constrains the total topological charge or winding number to be zero. In a nematic LC, trefoil particle knots with either type of surface anchoring tend to orient with their corresponding torus plane orthogonal to $\mathbf{n}_0$ in their ground state (Figure 5a,b) where the total elastic energy is the lowest but are also found in metastable orientations such as those with the torus plane parallel to $\mathbf{n}_0$. Boojums induced by a trefoil knot with planar anchoring are usually located at the particle's surface where the surface normal is parallel to $\mathbf{n}_0$ (Figure 5a,c). A total number of 12 boojums can be found in such cases for trefoil-knot-shaped particles, with half of them having the winding number of +1 and the other half −1, adding to the net total charge of zero to satisfy the topological constraint. Similarly, a pentafoil knot with planar anchoring in the stable orientation state will induce 20 such self-compensating defects (Figure 5d), or in other words, four times the number of turns the knot string winding around the circular axis of the corresponding torus. For trefoil knots with homeotropic surface anchoring, however, the particles usually induce closed disclination loops (Figure 5e). As an example, for the case of the particle's torus plane perpendicular to $\mathbf{n}_0$, one observes two defect loops that are both trefoil knots tracing along the tube of the knotted trefoil-shaped particle and linked with each other and with the particle knot, forming a three-component link of knots (inset of Figure 5e). Knotted disclination loops can also be induced by multiple topologically simple spherical particles in twisted LCs where complex configurations of closed disclination loops are promoted by chiral symmetry breaking [47,48].

Colloidal particles consisting of separate but linked components provide another way of generating complex topological defects, such as multi-component links. When suspended in LCs, these multi-component linked yet disconnected colloidal surfaces result in elastic coupling of the linked component through director deformations and topological defects that they share in their tubular surroundings. Martinez et al. fabricated Hopf and Salomon links with the linking numbers Lk = ±1 and ±2, respectively (Figure 5f–l) [49]. The linking number is a topological invariant representing the number of times that each individual constituting loop winds around the other. The topological constraint requires a net zero of the total winding number or hedgehog charge given that the Euler characteristic is $\chi = 0$ for all colloidal links studied. For Hopf links with planar surface anchoring, the most commonly observed director configuration contains eight boojums on the particle surface, with the two constituting colloidal unknot laying close to each other and their planes crossing (Figure 5f,h). The minimization of elastic energies associated with the director distortion and topological defects holds the linked components in their equilibrium positions despite thermal fluctuations. The elastic coupling between the constituting loops can be characterized by tracking their positional and angular diffusion (Figure 5g) and is found to be Hookean-like with the interaction force in the pN range. For Hopf links with homeotropic surface anchoring, the elastic coupling manifests itself in configurations where the induced singular defect lines jump from one component to the other, serving as an elastic string holding the two components together (Figure 5i,j). In the more complex cases of Salomon links, twice as many surface boojums tend to be induced compared with their Hopf counterparts, while pairs of individual disclination loops can be seen following each of the linked components with homeotropic surface anchoring (Figure 5k,l). However, many metastable particle orientations and director field configurations can be observed for linked particles of any given linking number and surface anchoring. A non-exhaustive survey presented in Ref. [49] shows that colloidal linked rings and accompanying singular defect loops can form many topologically distinct multi-component systems even with the same number of unknots and linking numbers.

The colloidal systems of links and knots in LCs described above exemplify the emergent behaviors and complexities of topological particles and director distortions even though topological constraints for particles of Euler characteristic $\chi = 0$ do not require the presence of topological defects. Importantly, the disclination loops and knots would not be even topologically stable in a polar system, so the fact that such a large variety of multi-component links and knots of defects can be found demonstrates the complexity of the interplay between topologies of surfaces and nonpolar molecular alignment fields.

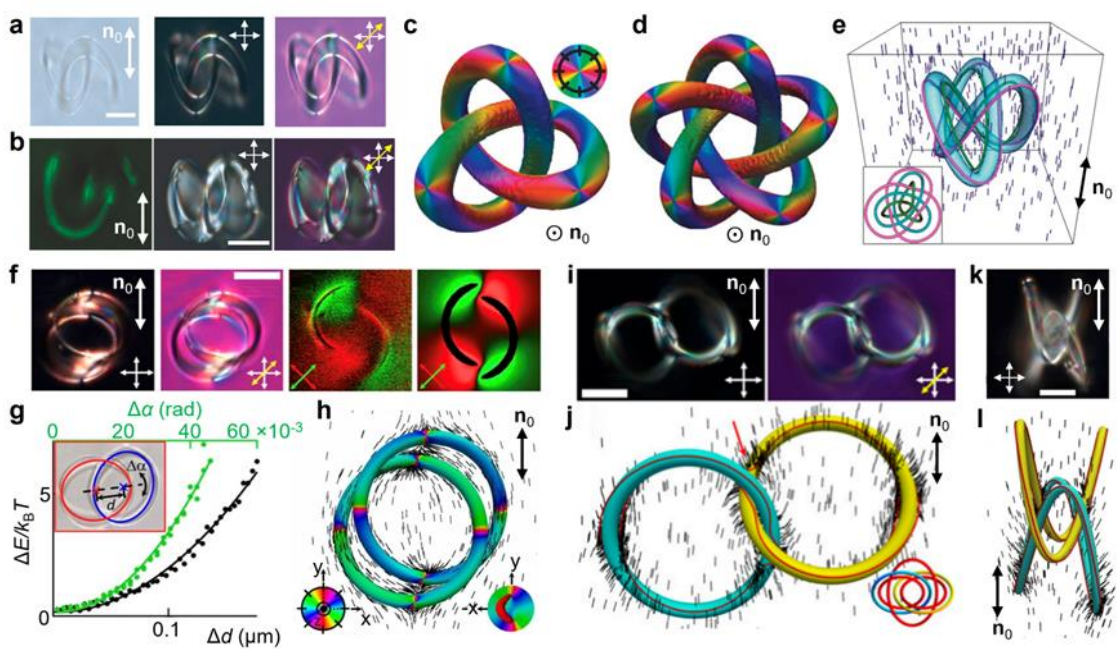

**Figure 5.** Knotted (**a**–**e**) and linked (**f**–**l**) colloids in LCs. (**a**,**b**) Optical micrographs of a trefoil knot with planar (**a**) and homeotropic (**b**) surface anchoring suspended in LCs. Crossed white double arrows indicate the direction of the polarizer and the analyzer and the yellow double arrow indicates the direction of the slow axis of a 530 nm retardation plate. (**c**,**d**) Numerically simulated director field distributions on the surface of trefoil (**c**) and pentafoil (**d**) knots with planar surface anchoring. The color represents the orientation of the surface director projected to the plane orthogonal to **n**$_0$; the color scheme is shown as the inset of (**c**). The far-field director **n**$_0$ is perpendicular to the sample plane as marked by the dot in a circle. (**e**) Schematic of defect lines (represented by green and magenta lines) induced by and entwined with a trefoil knot of homeotropic anchoring obtained from numerical simulation. (**f**) Polarizing, fluorescence, and simulated micrographs of linked colloidal rings with tangential surface anchoring suspended in LCs. The green and red double arrows represent the excitation polarization for fluorescence imaging. (**g**) Elastic interaction energy vs. deviation from the equilibrium center-to-center separation $\Delta d$ (black symbols) and orientation $\Delta \alpha$ (green symbols) as defined in the inset. (**h**) Numerically simulated director field distributions on the surface of a Hopf link at the position in (**f**). The color represents the director orientation as defined in the inset. (**i**, **k**) Polarizing micrographs of Hopf (**i**) and Salomon (**k**) links with homeotropic surface anchoring suspended in LCs. (**j**,**l**) corresponding simulated director field distributions and defect field (represented by the red lines) induced by the links. The red arrow in (**j**) points at the location where the disclination line jumps from one colloidal loop to the other. The double arrows marked **n**$_0$ represent the far-field director. Adapted from Refs. [46,49].

## 6. Conclusions and Future Perspectives

A central topic in colloidal science is to investigate the emergent behavior of constituting colloidal particles as they interact and self-assemble. They are ideal candidates for modeling the interaction and dynamics of atomic matter, such as self-assembly, crystallization, phase transition, etc., and even envisioning artificial forms of matter and metamaterials that go well beyond the complexity of natural systems [50–53]. Compared to the isotropic colloidal spheres and solvents commonly used in colloidal crystals, LC colloids provide new forms of long-range and anisotropic interactions mediated by the LC elasticity through director field deformation [5]. The examples described in this review have shown that varying the morphological characteristics of constituting particles such as symmetry and topology leads to a diversity of director configurations and colloidal behaviors. These colloidal "big atoms" with features of chiral symmetry, nontrivial topology, as well as knotted and linked components may readily contribute to the development of colloidal superstructures and exceed the diversity of atomic lattices. In addition, solitonic structures in cholesteric LCs involving topological defects and knotted and linked director configurations can also serve as building blocks of colloidal matter [6]. Although this review focused on nematic colloidal systems, other types of LC phases with additional partial positional degrees of ordering and chiral superstructures, such as cholesteric, columnar, and smectic LCs, may provide further symmetry-breaking director distributions and solvent-mediated interactions for LC colloids. These more complex LC field configurations add to the toolbox accessible for exploring fundamental mathematical and physical questions as well as developing novel functional materials with emergent properties and practical utility.

Historically, Lord Kelvin considered a model of atoms where different chemical elements in the periodic table would correspond to different knotted vortices [6]. Even though this model turned out to be incorrect, curiously, within the "colloidal atom" paradigm involving nematic colloids and vortices shaped as knots, researchers can now create modern-day analogs of Kelvin's topologically protected particles, thus opening doors to self-assembly of mesostructured materials with topology- and chirality-based design principles.

Future directions to be pursued as essential for the field of LC colloids may include the (i) large-scale self-assembly of low-symmetry and complex-topology colloidal particles; (ii) design and demonstration of out-of-equilibrium processes including stimuli-responsive dynamic rotational and translational behavior and the ensuing collective and emergent dynamic patterns [44,54]; (iii) design of colloidal composite materials of novel viscoelastic, electromagnetic, photonic, etc., properties; and (iv) use of ensuing novel composite materials to control transmission and directional scattering of light, as needed for applications in smart windows, displays, and electro–optic devices should be explored [55]. Like how knotted disclination lines guide the formation of crystal-lattice-like structures of colloidal spheres in chiral LCs [4,6], colloidal knots or other chiral and topological particles as well as LC solitonic structures may assemble into organization patterns not accessible to their atomic counterparts. Such novel assemblies may open possibilities to chiral, topological, and knotted colloidal metamatter with pre-designed physical properties.

The topology- and geometry-guided colloidal self-assembly approach combines soft and hard materials at the mesoscale in order to develop solids and fluids that exhibit emergent physical behavior not encountered in conventional material systems, with the inverse design and pre-engineering capabilities. Future research will need to focus on understanding and the control of material behavior arising from highly controlled mesoscale interpenetration of ordered solid and liquid crystalline building blocks, with the composites exhibiting solid, liquid, or mesophase behaviors controlled by weak stimuli. The design and realization of desired physical properties in these composites can be based on the following strategies: (1) templating the mesoscale organization of quantum and plasmonic nanoparticles by mesoporous structures enabled by lyotropic liquid crystals to harness quantum mechanical and plasmon–exciton interactions; (2) coupling between topological structures of continuous and singular field configurations in the solid and soft matter mesoscale subsystems; (3) discovery of low-symmetry fluids in hybrid molecular-colloidal systems enriched by exploiting particle's geometric shapes, surface charging and chirality; and (4) emergent effects in magnetic, colloidal, and topological solitonic spin ice systems arising in soft-hard matter systems. Electrical realignment of the LC component of the composites may in the future allow for the rearrangement and reorientation of anisotropic nanoparticles, leading to unprecedented control over self-assembled nanostructures and to dramatic changes in the material emergent behavior and properties. The LC colloidal research transcends traditional disciplinary boundaries of physics, topology, chemistry, engineering, and materials science and will advance our knowledge of mechanisms of the nanoscale self-organization phenomena of importance to achieving many scientific and technological goals of renewable energy research. Since our abilities to harvest, store, efficiently use, and convert energy among different forms are highly dependent on available materials and their properties, this research may lead to cheaper and more efficient renewable energy technologies, a new breed of energy-efficient information displays and electro–optic consumer devices, as well as a fertile ground for new basic science, with transformative impacts.

**Author Contributions:** Writing—original draft preparation, Y.Y.; writing—review and editing, Y.Y. and I.I.S. All authors have read and agreed to the published version of the manuscript.

**Funding:** Y.Y. acknowledges the financial support from the Japan Society for the Promotion of Science (JSPS KAKENHI grant number JP24K23088) when writing the manuscript. I.I.S. acknowledges the financial support of the fundamental research on colloids (the U.S. Department of Energy, Office of Basic Energy Sciences, Division of Materials Sciences and Engineering, under Award ER46921, contract DE-SC0019293 with the University of Colorado in Boulder).

**Acknowledgments:** I.I.S. acknowledges the hospitality of the International Institute for Sustainability with Knotted Chiral Meta Matter (WPI-SKCM$^2$) in Japan while on a sabbatical stay, as well as the hospitality of the Kavli Institute for Theoretical Physics in Santa Barbara while participating in the "Nanoparticle Assemblies" KITP program, when he was partially working on this review.

**Conflicts of Interest:** The authors declare no conflict of interest.